\documentclass[conference]{IEEEtran}

\usepackage{cite}

\ifCLASSINFOpdf
   \usepackage[pdftex]{graphicx}
   \graphicspath{{../pdf/}{../jpeg/}}

\else
   \usepackage[dvips]{graphicx}
   \graphicspath{{../eps/}}

\fi

\usepackage{amsmath}

\usepackage{algorithmic}

\usepackage{array}

\ifCLASSOPTIONcompsoc
  \usepackage[caption=false,font=normalsize,labelfont=sf,textfont=sf]{subfig}
\else
  \usepackage[caption=false,font=footnotesize]{subfig}
\fi

\usepackage{ulem}
\usepackage{color}
\usepackage[dvipsnames]{xcolor}

\usepackage{cite}

\usepackage{flushend}

\begin{document}

\newcommand{\vj}[2]{{\textcolor{magenta}{\sout{#1}}}\textcolor{cyan}{#2}}

\title{Personalized Brain-Computer Interface Models \\ for Motor Rehabilitation}

\author{\IEEEauthorblockN{Anastasia-Atalanti Mastakouri\IEEEauthorrefmark{1},
Sebastian Weichwald\IEEEauthorrefmark{1},
Ozan \"Ozdenizci\IEEEauthorrefmark{2},
Timm Meyer\IEEEauthorrefmark{1}, \\
Bernhard Sch\"olkopf\IEEEauthorrefmark{1} and
Moritz Grosse-Wentrup\IEEEauthorrefmark{1}}
\IEEEauthorblockA{\IEEEauthorrefmark{1}Empirical Inference Department, Max Planck Institute for Intelligent Systems 72076 {T\"ubingen}, Germany\\ Email: \{atalanti.mastakouri, sweichwald, timm.meyer, bs, moritzgw\}@tue.mpg.de}
\IEEEauthorblockA{\IEEEauthorrefmark{2}Faculty of Engineering and Natural Sciences, Sabanc{\i} University, Istanbul, Turkey\\ Email: oozdenizci@sabanciuniv.edu}}

\maketitle

\begin{abstract}

We propose to fuse two currently separate research lines on novel therapies for stroke rehabilitation: brain-computer interface (BCI) training and transcranial electrical stimulation (TES). Specifically, we show that BCI technology can be used to learn personalized decoding models that relate the global configuration of brain rhythms in individual subjects (as measured by EEG) to their motor performance during 3D reaching movements. We demonstrate that our models capture substantial across-subject heterogeneity, and argue that this heterogeneity is a likely cause of limited effect sizes observed in TES for enhancing motor performance. We conclude by discussing how our personalized models can be used to derive optimal TES parameters, e.g., stimulation site and frequency, for individual patients.
\end{abstract}


\IEEEpeerreviewmaketitle

\section{Introduction}

Motor deficits are one of the most common outcomes of stroke. According to the World Health Organization, 15 million people worldwide suffer a stroke each year. Of these, five million are permanently disabled. For this third, upper limb weakness and loss of hand function are among the most devastating types of disabilities, which affect the quality of their daily life \cite{who}. Despite a wide range of rehabilitation therapies, including medication treatment \cite{Walker}, conventional physiotherapy \cite{Green}, and robot physiotherapy \cite{Lum}, only approximately 20\% of patients achieve some form of functional recovery in the first six months \cite{Kwakkel, Nakayama}.

Current research on novel therapies includes neurofeedback training based on brain-computer interface (BCI) technology and transcranial electrical stimulation (TES). The former approach attempts to support cortical reorganization by providing haptic feedback with a robotic exoskeleton that is congruent to movement attempts, as decoded in real-time from neuroimaging data \cite{MGW1,Gomez}. The latter type of research aims to reorganize cortical networks in a way that supports motor performance, because post-stroke alterations of cortical networks have been found to correlate with the severity of motor deficits \cite{ANA:ANA21810,ANA:ANA21228}. While initial evidence suggested that both approaches, BCI-based training \cite{ANA:ANA23879} and TES \cite{doi:10.1093/brain/awh369}, have a positive impact, the significance of these results over conventional physiotherapy was not always achieved by different studies \cite{doi:10.1177/1550059414522229}, \cite{10.3389/fneng.2014.00030}, \cite{PMID:22964028}. 

One potential explanation for the difficulty to replicate the initially promising findings is the heterogeneity of stroke patients. Different locations of stroke-induced structural changes are likely to result in substantial across-patient variance in the functional reorganization of cortical networks. As a result, not all patients may benefit from the same neurofeedback or stimulation protocol. We thus propose to fuse these two research themes and use BCI technology to learn personalized models that relate the configuration of cortical networks to each patient's motor deficits. These personalized models may then be used to predict which TES parameters, e.g., spatial location and frequency band, optimally support rehabilitation in each individual patient.

In this study, we address the first step towards personalized TES for stroke rehabilitation. Using a transfer learning framework developed in our group \cite{TL}, we show how to create personalized decoding models that relate the EEG of healthy subjects during a 3D reaching task to their motor performance in individual trials. We further demonstrate that the resulting decoding models capture substantial across-subject heterogeneity, thereby providing empirical support for the need to personalize models. We conclude by reviewing our findings in the light of TES studies to improve motor performance in healthy subjects, and discuss how personalized TES parameters may be derived from our models.

\section{Methods}
\subsection{Subjects}

Twenty six healthy male subjects (mean age of 28.3 years with a standard deviation of 7.6 years) participated in this study, all of which were naive to the task and indicated that they are right-handed.
After a detailed explanation of the experiment, each subject gave informed consent in agreement with guidelines set by the ethics committee of the Max Planck Society which approved this study.

\subsection{Experimental Set-up}

The experimental set-up of this study consists of the following parts:

\subsubsection{A motion capture system}
We use the Impulse X2 Motion Capture System (PhaseSpace, San Leandro, CA, U.S.) which captures the $x,y,z$-coordinates of the subject's right arm position at a rate of 960 Hz.
For this, subjects are wearing a sleeve on their right arm equipped with infrared LEDs and the system's four infrared cameras are placed around them.

\subsubsection{Visual feedback screen}
During the experiment subjects are seated approximately 1.5 meters in front of a feedback screen.
The arm position, as constantly tracked by the motion capture system, is represented as a striped sphere on the screen.
The sphere is designed with a 3D stripe pattern and a visible shadow in order to facilitate the adaptation of the subject to the virtual 3D space on the screen.

\subsubsection{An EEG system}
We acquire the electroencephalogram (EEG) using an active 121-channel cap and a BrainAmp DC amplifier (BrainProducts, Gilching, Germany).
The sampling frequency is 500 Hz and the electrodes are positioned according to the 10-5 system for high-resolution EEG \cite{Oostenveld2001713}.
The reference electrode is placed at the TPP9h location.

\subsection{Experimental Paradigm}

The experimental paradigm is implemented using BCPY2000, an extended Python version of BCI2000 \cite{bci2000}. The experimental phases are described subsequently.

\subsubsection{Calibration phase}
At the beginning of each experiment, the subjects are instructed to place their arm in a comfortable position next to the leg. This position is defined as the ``starting position'' of the sphere on the screen.
Afterwards, the subjects are instructed to move the arm around in the space while always remaining seated and focusing on a fixation cross on the screen, in order to explore the area of comfortable movements for every subject.
During this ``exploration'' period, possible targets inside the limits of the subject's reaching area are computed.

\subsubsection{Resting phase}
There are five minutes of baseline recording, during which the participant is asked to focus on a fixation cross shown on the screen without moving.

\subsubsection{Trial phase}
\begin{figure*}[!t]
	\centering
	\includegraphics[width=.9\textwidth,keepaspectratio]{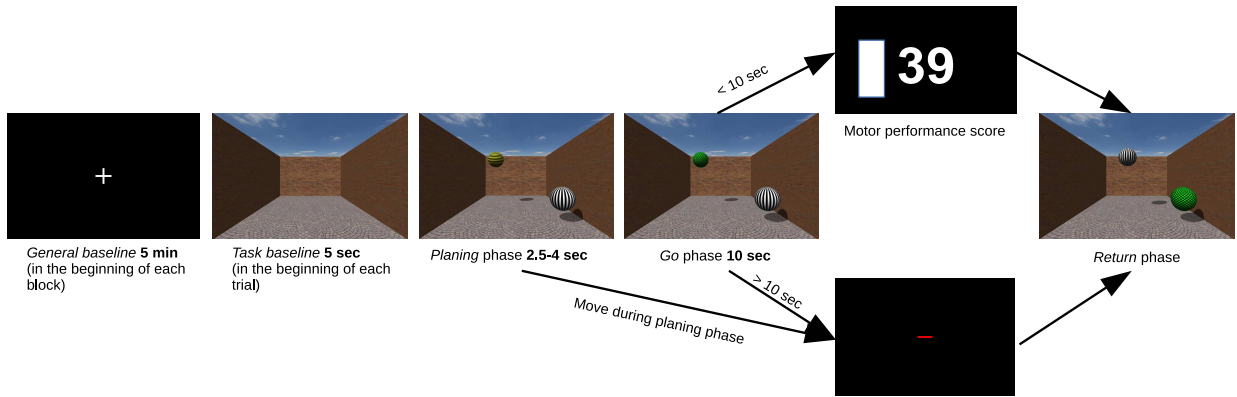}
	\caption{Trial sequence of the visuo-motor reaching task.}
	\label{fig_1}
\end{figure*}

The trial phase consists of two blocks of 50 trials where each block is followed by a five minute resting state recording.
In the following we describe the trial sequence which is also depicted in Fig.~\ref{fig_1}.
Each trial begins with a ``task baseline'' of 5 seconds, during which subjects are asked to rest and no sphere is shown on the screen.
During the following ``planning'' phase (duration uniformly chosen between 2.5--4 seconds) one white and one yellow patterned sphere is shown on the screen, the former reflecting the subject's arm position, the latter showing the randomly chosen target position for the next reaching movement.
Subjects are asked to plan the next reaching movement but not yet move.
Once the target sphere turns green, the ``go'' phase begins and this is the signal for the subject to move the arm and try to reach the target position.
A trial is considered ``failed'', and a black screen with red bar is shown, if subjects move more than 4 cm during the ``planning'' phase or if the target is not reached (overlapping of the end-effector sphere and the target not reduced below 3.5 cm) within the 10 second ``go'' phase.
Otherwise, the trial is considered successful and subjects receive feedback about their motor performance score (cf.\ Section~\ref{subsec:narj}).
Afterwards, the ``return'' phase starts, during which the subjects return, without time constrains, to the ``starting position'' which is now depicted as a green (target) sphere; once the white sphere representing her arm position overlaps 1 cm with the green sphere, the trial is considered completed.

\subsection{Index of Motor Performance}\label{subsec:narj}
In this study, we use the normalised averaged rectified jerk (NARJ) \cite{PMID:12488089} as an index of motor performance.
It reflects the \textit{smoothness} of a movement and was shown to correlate with the Fugl-Meyer Assessment of Motor Recovery after Stroke (FMA) \cite{6095107}.

We compute the NARJ for each movement from the jerk values $\operatorname{Jerk}_{\cdot,t}$, the second derivative of the velocity, at each time step $t$ in each of the three dimensions $x,y,z$ tracked by the motion capture system as follows
\begin{equation*}
\operatorname{NARJ} = T^3 \frac{1}{T} \sum_t \sqrt{\operatorname{Jerk}_{x,t}^2 + \operatorname{Jerk}_{y,t}^2 + \operatorname{Jerk}_{z,t}^2}
\end{equation*}
where $T$ is the duration of the reaching movement.
What we show subjects as feedback on the screen is the inverted NARJ fitted in a range between 0 and 100, so that a higher score can be interpreted by subjects as reflecting a ``better'' movement.

\subsection{EEG Analysis}

\subsubsection{Preprocessing}
Removing the \textit{``failed''} trials results in 89--98 trials per subject.
We restrict our analysis to the 118 EEG channels that were consistently recorded with high quality for all subjects and removed the ones that were noisy in at least one of the recordings.
These were then rereferenced to common average reference.
We keep the EEG data in the time window 7.5--17.5 seconds of each trial, where 7.5 corresponds to the earliest possible start of the ``go'' and 10 seconds is its maximum duration.
In order to attenuate non-cortical artifacts, we perform an independent component analysis (ICA) and only reproject those independent components that, by visual inspection of the topographies and source frequency spectra, correspond to cortical sources (cf.\ Section~2.3 of \cite{grosse2012high} for a description of this procedure).

\subsubsection{Feature computation}
For each trial and channel we compute the log-bandpower in the following five frequency bands:
\textit{delta} ($\delta$, 1--4 Hz), \textit{theta} ($\theta$, 4--8 Hz), \textit{alpha} ($\alpha$, 8--13 Hz), beta ($\beta$, 13--30 Hz) and \textit{high gamma} ($\gamma$, 60--90 Hz) (the 30--60 Hz band is excluded due to the 50 Hz power line noise).
This results in a feature array for each subject of the form 118 channels $\times$ 5 logarithmic bandpowers $\times$ number of trials.

\subsubsection{Transfer learning regression}
We want to predict, for each trial individually, the logarithmic NARJ value from the log-bandpower features of the ``go'' phase of that trial.
We adapt the transfer learning algorithm presented in \cite{TL} in order to perform linear regression.
This enables us to leverage the data of 25 subjects when training a model for the 26$^{\text{th}}$ subject.
In particular, for every subject $s$ we train a predictive model with features from all the trials of the remaining 25 subjects.
This is the \textit{prior} model of subject $s$.
We then update the prior model's weights with the data from the first 20 trials of subject $s$.
We call this model the \textit{updated} or \textit{personalized} model for subject $s$.
This \textit{personalized} model is then used to predict the remaining trials of this subject. We use leave one subject out cross validation in order to evaluate our model.

\subsection{Statistical Tests}
\label{subsec:stats}
In order to evaluate the predictive power of our models, we first assess their ability to correctly predict the average NARJ value in the final 50 trials of each subject.
For this, we compute the across-subject correlation coefficient between the predicted average NARJ values and the observed ones.
To estimate the p-value under the null-hypothesis that the predicted and observed average NARJ values are uncorrelated, we permute the subject-order of the predicted average NARJ values $10^4$ times and compute the instances in which the modulus of the resulting correlation coefficient exceeds the modulus of the correlation coefficient with the subject-order intact.

We quantify the ability to predict the NARJ value of individual trials over the course of the experiment by the magnitude square coherence between the predicted and the observed NARJ values for each subject.
Coherence is commonly used to estimate the power transfer between input and output of a linear system; it estimates the extent to which one signal can be predicted from another by an optimum linear least squares function \cite{Bendat}.
We then randomly permute, within subjects, the trial-order of the predicted NARJ values $10^4$ times and compare the resulting magnitude square coherence with the magnitude square coherence for the correct trial-order.
This yields a p-value for each subject assessing the magnitude square coherence.

To extend this to a group-level test, we use the fact that, by definition, p-values are drawn from a standard uniform distribution if the null-hypothesis is true.
To quantify the deviation of the empirical cumulative distribution function (CDF) of above p-values from the CDF of a standard uniform distribution we create one hundred equally sized bins between zero and one and sum, across all bins, the absolute differences between the empirically observed CDF and the one generated by drawing the
same number of samples from a standard uniform distribution.
Sampling this test statistic $10^3$ times gives us a p-value reflecting how likely it is that the subjects' p-values are drawn from a standard uniform distribution.

\section{Results}
\subsection{Adaptation of Motor Performance over Time}
\begin{figure}[!t]
	\center\
	\includegraphics[width=\linewidth,keepaspectratio]{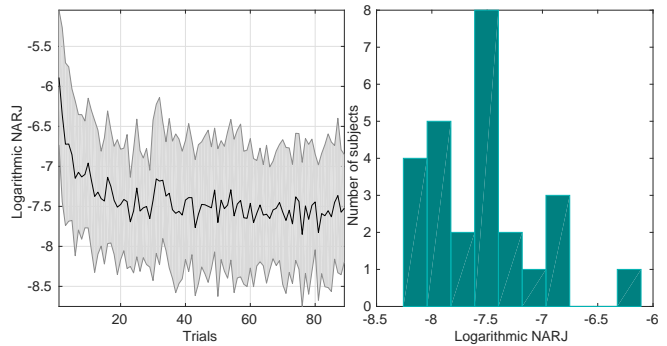}
	\caption{Left: Motor performance over time (mean $\pm$ one standard deviation). Right: Histogram of performance in the end of the experiment (averaged over last 50 trials).}
	\label{fig_2}
\end{figure}

The left column of Fig.\ref{fig_2} displays the mean and standard deviation of the logarithmic NARJ values across subjects for the first 89 trials (minimum number of trials available across subjects). There is a strong adaptation period during the first 20 trials. After roughly 50 trials, the mean NARJ values have almost converged to their final value. The distribution of final movement smoothness across subjects is shown in the right column of Fig. \ref{fig_2} (averaged over last 50 trials), exhibiting a substantial heterogeneity of subjects' final performance.

\subsection{Model Validation}
\subsubsection{Prediction of subjects' final mean motor performances}

\begin{figure}[!t]
	\center\
	\includegraphics[width=\linewidth,keepaspectratio]{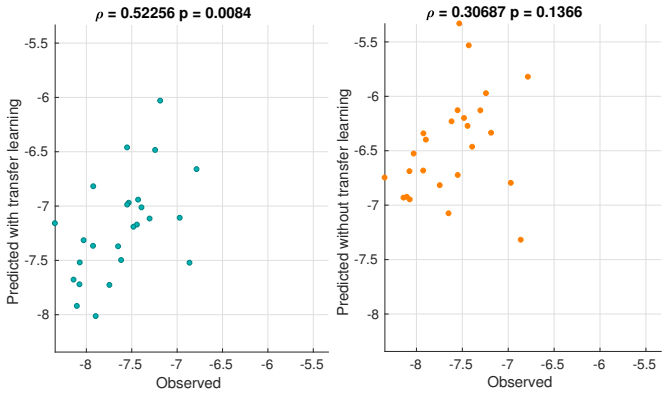}
	\hfil
	\caption{Predicted and observed average logarithmic NARJ of the last 50 trials for the 26 subjects with (left) and without (right) the use of transfer learning.}
	\label{fig_6}
\end{figure}

Figure \ref{fig_6} shows, for each subject, the observed versus the predicted average NARJ values in the final 50 trials both for the personalized (left column) and prior model (right column).
Only the updated model exhibits a significant correlation between model predictions and observed true values ($\rho=0.52,\ p = 0.008$) while for the predictions of the prior model there is not sufficient evidence to reject the null-hypothesis of chance-level performance ($\rho=0.31,\ p=0.139$).

\subsubsection{Prediction of motor performance in individual trials}

\begin{figure}[!t]
	\centering
	\includegraphics[width=.8\linewidth,keepaspectratio]{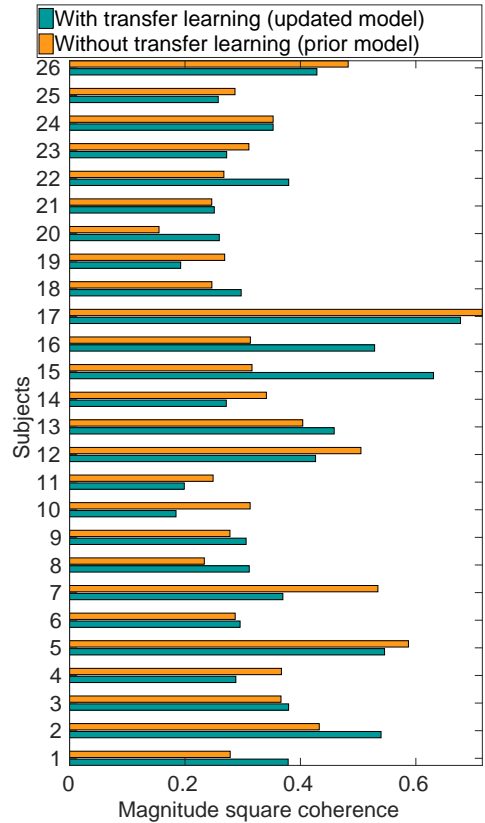}
	\caption{Magnitude-squared coherence between observed and predicted NARJ values for each subject.}
	\label{fig_coherence}
	\hfil
\end{figure}

\begin{figure*}[!t]

	\center
	\includegraphics[width=.9\textwidth]{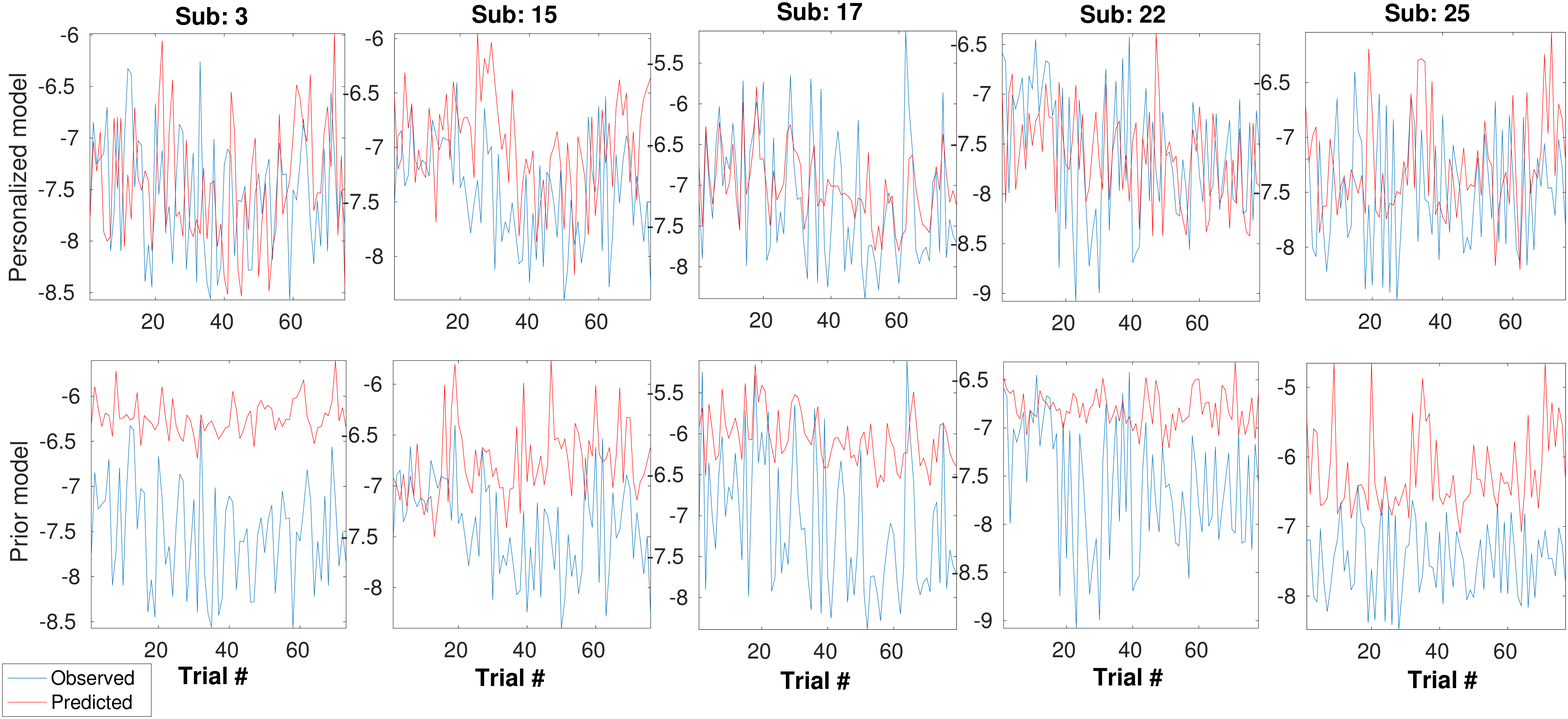}
	\hfill
	\caption{Observed (blue) and predicted (red) NARJ values across trials for five representative subjects. Top row: personalized model. Bottom row: prior model.}
	\label{fig_trial_pred}
\end{figure*}

The ability of our models to predict movement smoothness for individual trials is assessed by the magnitude square coherence (cf.\ Section~\ref{subsec:stats}) as shown in Fig.~\ref{fig_coherence} for each subject and model type (updated and prior).
While both models achieve a similar mean coherence across subjects---$0.36$ and $0.35$ for the updated and prior model respectively---the comparison of the distribution of $p$-values across subjects with the CDF of a standard uniform distribution reveals a notable difference.
While the $p$-value under the null-hypothesis of standard uniformly distributed subject p-values is marginally significant for the updated model ($p = 0.06$), the prior model returns a $p$-value of $0.63$. Figure~\ref{fig_trial_pred} shows observed and predicted NARJ values across trials for five representative subjects for the updated model (top row) and the prior model (bottom row).
It is apparent from this plot that only the updated model captures meaningful variations in movement smoothness across trials.

\subsection{Model Interpretation}
\begin{figure}[!t]
	\center
	\includegraphics[width=.95\linewidth,keepaspectratio]{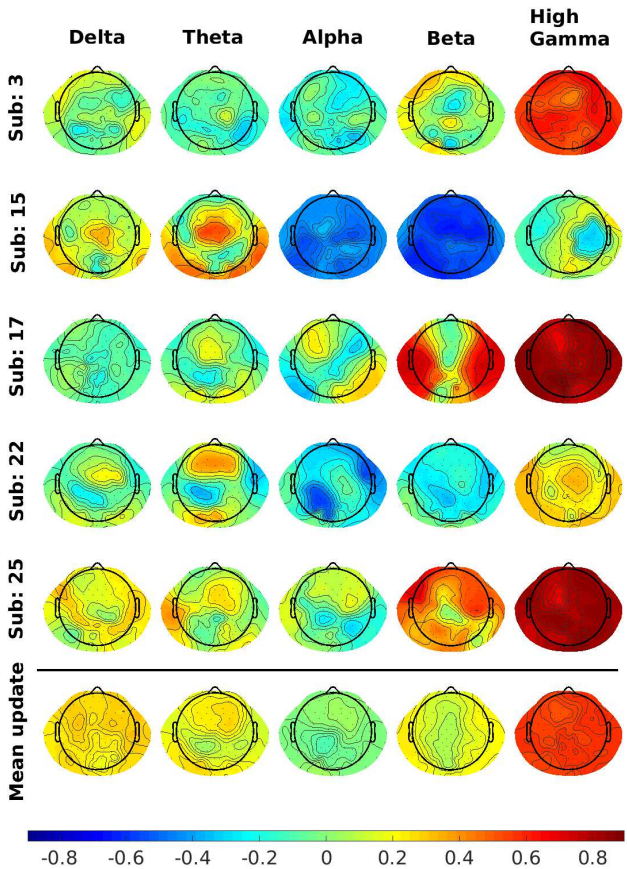}
	\caption{Personalized encoding models, i.\,e.\ feature correlation with predicted NARJ values, for five representative subjects (first five rows) across the five frequency bands (five columns).
		The last row shows the mean encoding model averaged over all subjects.} \label{fig_7}
\end{figure}

To gain a better understanding of the cortical processes used for prediction, we compute the correlation coefficients between the personalized models' predictions and the individual electrode bandpower features. That is, we quantify how much each channels' bandpower contributes to the prediction, i.\,e.\ we essentially obtain an encoding model from our decoding model \cite{Haufe201496}.
The resulting \textit{encoding} topographies (5 representative subjects and mean topography) are shown in Figure \ref{fig_7}. Red/blue color indicates a positive/negative correlation between electrode bandpower and the logarithmic NARJ, i.\,e., increased bandpower at blue colored electrodes is associated with smoother movements.
We note, first, that there is a qualitative difference between the average model and the personalized encoding models, and, second, that the personalized models exhibit substantial heterogeneity.
Strongest correlations---but with inconsistent signs---are observed in the alpha, beta and high gamma range, while correlations in the delta and theta range are comparably small across subjects.

\section{Discussion}

\subsection{Transfer Learning for Personalized Models}

We have shown in this work how to build personalized models that relate the global configuration of EEG rhythms to motor performance.
This enables us to cope with the heterogeneity of motor performance across subjects as well as extend previous work \cite{MeyerPZSG2013} to the harder task of single-trial prediction \cite{tanger}.
A crucial feature of our approach is to employ a transfer learning framework.
Because of the high dimensionality (590-D) of our feature space, building personalized models based on subject-specific training data would require several hundreds of training trials, resulting in a calibration time of several hours.
Using the transfer learning framework enabled us to learn each personalized model from only 20 trials of that very subject.
Because our long-term goal is to make personalized predictions on optimal TES parameters for stroke rehabilitation, it is essential that the calibration of our models is fast enough to be applied in a clinical setting.

\subsection{TES and Model Heterogeneity}

While most TES motor studies consistently focus on the contralateral motor cortex M1, their individual findings are inconsistent with one another inasmuch as they evidence effects of contradicting quality for the different frequency bands:
Some studies report an inhibiting effect of transcranial alternating current stimulation (tACS) at 20 Hz over the contralateral motor cortex on motor performance, but no significant effect of stimulation in the gamma frequency range \cite{Pogosyan20091637,beta,TJP:TJP4253}; others, seemingly contradicting, describe significant effects in the gamma range and do not find significant evidence for inhibiting effects of stimulation in the beta range \cite{Moisa12053}.
In frontoparietal areas, gamma oscillations were found to be correlated with reaction times in a motor task \cite{HBM:HBM20056}, in contrast to stimulation studies that found improvements in implicit motor learning only after applying 10 Hz AC but for neither 1, 15, 30 or 45 Hz \cite{Antal200897}.

In general, the heterogeneity in the organization of subjects' cortical networks may explain such inconsistent results.
Our findings further support this line of argument by evidencing a substantial heterogeneity amongst subjects:
The activity in the alpha, beta, and gamma range turns out to be sometimes negatively and sometimes positively correlated with motor performance.
That is, when using one stimulus protocol (location and frequency) for all subjects instead of personalized stimulation protocols, the differences between subjects may lead to inconsistent group-level results.
In line with previous findings, our models reveal the alpha, beta and (high) gamma frequency ranges as decisive for motor performance.
Our results indicate that a one-for-all stimulation approach is unlikely to consistently improve motor performance and emphasize the importance of personalized stimulus protocols focusing on the alpha, beta and gamma ranges.

\subsection{Predicting Optimal TES Parameters}

Decoding models as the ones trained in this study do not immediately reflect causal relationships and as such do not allow to directly read off optimal stimulation parameters for each subject \cite{Haufe201496,weichwald2015causal} (see specifically interpretation rules R3 and R4 in \cite{weichwald2015causal}).
While encoding models allow us to rule out EEG features that are not causal for motor performance (cf.\ interpretation rule R2), they cannot be used to identify the causes of a behavioral response (cf.\ interpretation rule R1).
In light of recent work that has demonstrated that the combination of both encoding and decoding models enables richer causal interpretations than any model alone \cite{weichwald2015causal,huth2016decoding,bach2017whole}, we argue that future research on stroke rehabilitation should leverage this approach and fuse TES and BCI approaches:
by comparing feature relevance in both encoding and decoding models the search space over stimulation parameters for TES may be reduced.
In particular, we can safely restrict our search on the features that are both important in the encoding and decoding as we are guaranteed to expect causes of motor performance, if at all, only amongst these (cf.\ interpretation rules R5-R8).
Thus, we argue that a decoding model that is able to sufficiently well predict single-trial motor performance is a necessary prerequisite for personalized stimulation protocols and thus view our work as the first step towards personalized BCI-TES based stroke rehabilitation.

\bibliographystyle{IEEEtran}


\end{document}